\documentstyle[proceedings]{crckapb}

\begin{opening}
\title{GRAVITATIONAL COLLAPSE AND FRAGMENTATION IN MOLECULAR CLOUDS
        WITH ADAPTIVE MESH REFINEMENT HYDRODYNAMICS}

\author{RICHARD I. KLEIN}
\author{ROBERT T. FISHER}
\author{CHRISTOPHER F. MCKEE}
\author{J. KELLY TRUELOVE}
\institute{Astronomy Department, University of California\\
                601 Campbell Hall\\
                Berkeley, Ca. 94720}

\end{opening}

\runningtitle{GRAVITATIONAL COLLAPSE AND FRAGMENTATION}

\newcommand{\beq}{\begin{equation}}
\newcommand{\bo}{\beta_\Omega}
\newcommand{\cm}{\;{\rm cm}}
\newcommand{\cs}{c_s}
\newcommand{\dt}{\Delta t}

\newcommand{\dx}{\Delta x}
\newcommand{\eeq}{\end{equation}}
\newcommand{\gcc}{\;{\rm g}\;{\rm cm}^{-3}}
\newcommand{\jmax}{J_{\rm max}}
\newcommand{\jl}{\lambda_J}
\newcommand{\kms}{\;{\rm km}\;{\rm s}^{-1}}

\newcommand{\rads}{\;{\rm rad}\;{\rm s}^{-1}}

\newcommand{\rcrit}{r_{\rm crit}}

\newcommand{\totenergy}{E}
\newcommand{\vecQ}{\mbox{\bf Q}}
\newcommand{\vecv}{\mbox{\bf v}}

\begin{document}

\begin{abstract}
We describe a powerful methodology for numerical solution of 3-D
self-gravitational hydrodynamics problems with extremely high resolution.
Our method utilizes the technique of local adaptive mesh refinement
(AMR),
employing multiple grids at multiple levels of resolution. These grids are
automatically and dynamically added and
removed  as necessary to maintain adequate resolution.
This technology allows for the solution of problems in a manner that 
is both more efficient and more versatile than other fixed and variable
resolution methods.
The application of AMR 
to simulate the collapse and fragmentation of a
molecular cloud, a key step in star formation, is discussed.  Such simulations
involve many orders of magnitude of
variation in length scale as fragments form.
In this paper we briefly describe the methodology and present an illustrative application for 
 nonisothermal cloud collapse.
We describe the numerical
Jeans condition, a criterion for stability of self-gravitational
hydrodynamics problems.
We show the first well-resolved nonisothermal evolutionary sequence beginning
with a perturbed dense molecular cloud core that leads
to the formation of a binary system consisting of protostellar cores
surrounded by distinct protostellar disks. The scale
of the disks, of order 100 AU, is consistent with observations
of gaseous disks surrounding single T-Tauri stars and
debris disks surrounding systems such as $\beta$ Pictoris.
\end{abstract}

RIK is also affiliated with Lawrence Livermore National Laboratories,
L-023, Livermore, California, 94550.
\vfill\eject


\section{INTRODUCTION}
 Star formation ultimately
involves the collapse of an interstellar molecular cloud core from an
initial density of $\rho \leq 10^{-19}\gcc$ and size $\simeq 10^{17} \cm$ to a final
young stellar object of density $\rho \geq 10^{-1}\gcc$ and size $\simeq 10^{11}
\cm$.
The collapse of a portion of a cloud
may lead to fragmentation, which is crucial to
establishing key parameters of binary stars: the distributions of mass
ratios, periods, and orbital eccentricities.  It may also
be essential to the
formation of larger groups of stars and to the determination of the
stellar initial mass function. 
However, this
enormous dynamic range
presents a formidable obstacle to obtaining an accurate
numerical solution, as the flow must remain well-resolved throughout
the evolution.
Fixed-resolution methods  
cannot be used to simulate such a 3-D collapse in a practical amount
of time using current computers.

In order to treat the enormous dynamic range presented by gravitational
collapse calculations and address the problems associated with alternative methods
such as smoothed particle hydrodynamics (SPH) and static nested-grid codes (\citeauthor{bur93}, \citeyear{bur93}), we have developed a 3-D self-gravitational
adaptive mesh refinement
(AMR) code. Specifically, SPH suffers from two major drawbacks. First, it is typically very diffusive, and has great difficulty tracking shocks and contact discontinuities accurately in multi-dimensional flows. Second, although it is an adaptive method, it is only adaptive in the sense that the local smoothing length $h$ tracks mass in a Lagrangian sense, i.e. in 3-D, $h \propto \rho^{-1/3}$. Furthermore, in the context of isothermal collapse, in which the Jeans length $\lambda_{J}$ scales as $\lambda_{J} \propto \rho^{-1/2}$, the effective resolution, specified by the ratio $\lambda_J / h$ varies as $\rho^{-1/6}$ -- {\it the effective resolution of the calculation is degraded as collapse proceeds, and if the collapse is not arrested first, the calculation will eventually become underresolved}. Given
that in SPH one cannot choose {\it a priori} which regions of the flow need further 
refinement, one's only recourse to increasing the resolution of a calculation is
to increase the total number of particles. Lastly, static nested-grid codes suffer from
the drawback of imposing 
a static set of grids from the beginning of the calculation, and hence are 
unable to resolve arbitrarily-located fragments. 

Our AMR method dynamically resizes and repositions
grids and inserts new, finer ones within them according to
adjustable refinement criteria (unlike the static 
nested grid scheme of \citeauthor
{bur93} \citeyear{bur93}).
We have applied our method to great effect;
prior to the addition of self-gravity, we have used AMR to study 
astrophysical problems
including the interaction of supernova blastwaves with interstellar
clouds
(\citeauthor{kle94b} \citeyear{kle94b}; \citeauthor{kle94a} \citeyear{kle94a};
\citeauthor{kle95} \citeyear{kle95}),
X-ray heated coronae and winds from accretion disks \cite{woo96},
and the collision of interstellar clouds \cite{kle98} with unprecedented
high resolution.

We present the methodology behind our 3-D 
AMR self-gravity code and describe application of our work
to the collapse and fragmentation of molecular clouds.  Our discussion closely
follows \citeauthor{tru98} \citeyear{tru98} and \citeauthor{klein98} \citeyear{klein98}.

\section{METHODOLOGY}

\subsection{Self-Gravitational Hydrodynamics}
\label{sec-methodology overview}

The basic governing equations of our model are the Euler equations
of hydrodynamics in 3-D, including
effects of self-gravitation,
and Poisson's equation for the gravitational potential.
We term these the gravitational hydrodynamics (GHD) equations.
The total non-gravitational
energy per unit mass, $\totenergy$,
is related to the internal energy per unit mass, $e$, by
$\totenergy=e+\frac{1}{2}\vecv^2$.
In turn, $P$ is defined by an equation of state. In our isothermal
work, we  
adopt the ideal gas law
$P=(\gamma-1)\rho e$.
In recent work (\citeauthor{fisher98} \citeyear{fisher98}) we have used
a nonisothermal EOS that combines isothermal and nonisothermal
components such that $P(\rho) = c^2_s \rho + K\rho^\gamma$
where $\gamma=5/3$ and $K$ is chosen so that the isothermal and
adiabatic components balance at a critical density determined
by detailed 1-D radiation-hydrodynamic calculations of
\citeauthor{masu98} (\citeyear{masu98}).  A typical critical density for this
transition is $5\times 10^{-14}\gcc$.
As written above, there are eight equations in eight
unknowns.  We assume periodic boundary conditions on these
equations.

Our code methodology consists of three components to efficiently
solve this system.  The first component is a
hyperbolic solver
that employs an implementation of
the Godunov method (see \citeauthor{col84} \citeyear{col84} and \citeauthor{lev92} \citeyear{lev92})
for solution of the Euler
equations of gas dynamics.
The
second major component of our code methodology is
an elliptic solver that utilizes an AMR multigrid method to solve
Poisson's equation.  These elements operate within the
third component, an adaptive mesh refinement framework.

\subsection{Hyperbolic Solver}
\label{sec-hyperbolic}

\subsubsection{Overview of Godunov Implementation}

To solve the full 3-D system of hydrodynamics
equations, we use an
operator-split method in which we solve
in each coordinate direction independently in a cyclic sequence.
In 3-D, this operator splitting can be written schematically as 
\beq
\vecQ^{n+2} = L_x L_y L_z [ L_z L_y L_x (\vecQ^n) ],
\eeq
where the superscript $n$ indicates the $n^{\rm th}$ time level,
$\vecQ^n$ indicates the state vector at the $n^{\rm th}$ time level, and
$L_i$ is the update operator in the $i^{\rm th}$ coordinate direction
based on the higher-order
Godunov method
described by \citeauthor{col84} (\citeyear {col84}) and \citeauthor{bel89}
(\citeyear {bel89}). 
That is, each three-dimensional update
step from $t^n$ to $t^{n+1}$
consists of 3 one-dimensional update steps, one per coordinate direction.
At the end of the cycle from $t^n$ to $t^{n+2}$,
the solution is second-order accurate.
The interested reader can refer to
\citeauthor{tru98} (\citeyear{tru98}) and
\citeauthor{klein98} (\citeyear{klein98}) for the details of the scheme. 

\subsubsection{AMR Considerations and AMR Multigrid}
When solving a hyperbolic 
system on a hierarchy of computational
grids, care must be taken that the solutions on
finer grids
are reflected on the neighboring and underlying coarser grids.
Our code uses {\em refluxing} and {\em averaging down}
procedures described in detail by \citeauthor{ber89} (\citeyear{ber89})
to update the solutions on these coarser grids
to account for solutions on the finer grids.
In refluxing, the solution in a coarse cell adjacent to a fine grid is
updated after the fine grids have been advanced in time 
by adding the effects of a differential flux acting over
$\Delta t$ at its face
adjacent to the refined region.
This differential flux is equal
to the difference between the flux at this
face as computed on the coarse grid
and the sum of the fluxes at this face as
computed on the fine grid.
The solution in a refined coarse cell
is simply overwritten by averaging down
the solutions on the
fine cells it contains.

In contrast to the hyperbolic solution, in which only neighboring grids
communicate at each time step,
the solution of Poisson's equation for the gravitational potential is a
much more tightly coupled process.
The multigrid method is a natural choice to use as an element of the
elliptic solver in an AMR calculation (\citeauthor{tru98} \citeyear{tru98}).

\subsection{AMR Framework and the Refinement Criterion}
\label{sec-Jeans condition}

As discussed above, the basic hydrodynamic method we use is a higher-order
extension of Godunov's method of a type discussed in \citeauthor{col84} (\citeyear{col84}).
 This algorithm is second-order accurate
for smooth flow problems, and has a robust and accurate treatment of
discontinuities \cite{kle94b}.
We further enhance the
 efficiency and accuracy of our calculations by using local adaptive mesh 
refinement.

The version of the algorithm used for the present work is rather 
elaborate; a detailed description is given in \citeauthor{ber89} \citeyear{ber89}.

	AMR contains five separate components.  The {\it error estimator}
is used 
to estimate local truncation error and will be described subsequently.  
The {\it grid 
generator} creates fine grid patches that cover the regions that need 
refinement.  {\it Data structure routines} manage the grid hierarchy allowing
access to the individual patches.  {\it Interpolation routines} initialize a
solution on a newly created fine grid and also provide the boundary 
conditions for integrating the fine grids.  {\it Flux correction} routines 
ensure conservation at grid interfaces by modifying the coarse grid 
solution for coarse cells that are adjacent to a fine grid.

  In AMR, the computational volume
consists of a hierarchical grid structure.  A base Level 0 grid fills
the computational volume, discretizing it on a rectangular grid
with a resolution of $\Delta x_0$ in each direction.
Multiple Level 1 grids of finer resolution $\Delta x_1=\Delta x_0/r_1$ may be
embedded within it, where $r_1=4$ is a typical choice.
Higher levels may be defined embedded in underlying levels in a similar fashion.
Grids at Level $L$ always span
an integral number of cells at Level $L-1$, i.e., partial cell refinement
is not permitted.  Furthermore, an interior
grid at Level $L$ is always nested within
a grid at Level $L-1$ such that there is a buffer region of Level $L-1$ cells
surrounding it.  In other words,
a grid at Level $L$ within a grid at Level $L-1$ never
shares a boundary with the Level $L-1$ grid, unless it shares a boundary
with the computational domain.

An integration step on this adaptive grid
proceeds as follows.  
We are using a single timestep for all grids,
determined from the most stringent timestep
over the entire physical domain. 
The updating of the data 
on the locally refined grid structure is organized around the grouping
 of cells into rectangular grid patches, each typically 
containing hundred to several thousand grid cells. 
The AMR code passes a grid patch to an integration routine which
returns the updated cell values as well as the edge-centered conservative
fluxes used in the update.
The overheads in both
 CPU and memory associated with the adaptive mesh structure have been
kept quite small, relative to irregular grid schemes.
Typically, 80\% - 90\% of the total execution time is spent advancing 
cells in time using the
hydrodynamic and elliptic solvers,
while the memory 
required is that needed to store two copies of the solution on all
of the grids. These overheads are low because they are determined 
by the number of rectangles into which the AMR solution has been 
divided; as opposed to being determined by the number of grid cells,
 as is the case with irregular grid adaptive algorithms. 

A key component of an AMR code is
the grid generator. In our code, this procedure is
broken into two steps.  In the first step, a specified property is
measured in each cell, and cells requiring refinement are flagged.
In the second step, the distribution of flagged cells 
is analyzed to determine the number, sizes,
and locations of grids to be inserted at the next finer level of
resolution.  These finer grids will always include every cell
that was flagged for refinement, but they may also include
additional cells that were not flagged.
The degree to which the refinement is concentrated in
the cells that require it is termed the {\em grid efficiency}.
The grid efficiency is minimal when the smallest
rectangular solid containing all flagged cells is refined.
The grid efficiency is maximal when
the only cells refined are those that were flagged.  

Our general process for deciding upon a suitable level of refinement depends
on a method for estimating the local truncation error of the
Godunov scheme; this error estimator determines where the solution accuracy
is insufficient.  We estimate truncation error using 
Richardson extrapolation; 
we take data on the 
grid where the error is being estimated and evolve two timesteps.  This
solution is then combined with the data on a grid spatially coarsened by
a factor of two and integrated for one timestep.  The local truncation error
can be shown to be proportional to the normed functional that is the difference
of these two operations. 

In the nonisothermal calculations presented
in this paper and in the isothermal calculations of 
\citeauthor{tru97} (\citeyear{tru97}),
refinement to Level 2 and
above is controlled by application of the Jeans condition.
This criterion can be augmented or replaced by others as necessary.
The Jeans condition used to make the cell-by-cell refinement decision
is a physically motivated resolution constraint
discussed by \citeauthor{tru97} (\citeyear{tru97}).
Jeans (\citeauthor{jea02} \citeyear{jea02} and \citeauthor{jea28} \citeyear{jea28})
analyzed the linearized equations of
1-D isothermal GHD
for a medium of infinite extent and found that
perturbations on scales larger than the Jeans length,
$\jl\equiv\left(\frac{\pi\cs^2}{G\rho}\right)^{1/2}$,
are unstable.
Thermal pressure cannot resist the self-gravity of
a perturbation larger than $\jl$, and runaway collapse results.
\citeauthor{tru97} (\citeyear{tru97}) showed that the errors
generated by numerical GHD solvers can act as
unstable perturbations to the flow.
In a simulation with variable resolution, cell-scale
errors introduced in
regions of coarser resolution can be advected to regions of
finer resolution, affording these errors the opportunity to grow.
The unstable collapse of numerical perturbations
can lead to substantial fragments, a
process termed {\it artificial fragmentation}, which
can be avoided if
$\jl$ is sufficiently resolved. Defining the Jeans number
resolution of $\jl$.  Defining the Jeans number
$J\equiv\frac{\dx}{\jl}$,
\citeauthor{tru97} (\citeyear{tru97}) found keeping $J\leq0.25$ avoided
artificial fragmentation in isothermal evolution of a
collapse spanning 7 decades of density, the approximate
range separating
typical molecular cloud cores from nonisothermal protostellar
fragments.  The constraint that
$\jl$ be resolved is termed the {\it Jeans condition}.
Although it has
been shown to be necessary only for isothermal evolution, it is reasonable to
expect that it is necessary (although not necessarily sufficient) for
nonisothermal collapse as well.

As a side effect of
confining cell-sized perturbations to a length scale
at which they are thermally damped, resolution of $\jl$ also
ensures that 
gradients developed in isothermal flow by gravity are well
resolved.  Formation
of structure on scales of $\jl$ and larger is a general
feature of
self-gravitational isothermal GHD flow since smaller fluctuations are damped but larger
ones collapse.  For example, in the self-similar solution for
isothermal cylindrical
collapse \cite{inu92}, the radial scale height of
the cylinder is $\jl$.
Lack of resolution of gradients within simulated flow
triggers the injection of artificial viscosity, which is generally
intended to be introduced in small amounts only for numerical stability
(in our Godunov scheme)
or shock mediation (in many common non-Godunov schemes).
Introducing excess amounts of artificial viscosity
renders the problem solved different
from the inviscid problem posed.
Continuous resolution of $\jl$, however,
keeps the flow inviscid and prevents artificial slowing of
gravitational collapse. It is important to note, however,
that the Jeans condition is a necessary but not,
in general, sufficient condition to ensure convergence.

The Jeans condition on $\dx$ fundamentally differs from the Courant
condition on $\dt$, although at first the two conditions might appear
analogous.
The Courant condition arises from a modal stability
analysis of finite difference equations derived from the
Euler partial differential equations (PDEs) of
hydrodynamics (see, e.g., \citeauthor{ric67} \citeyear{ric67}).
It is entirely a consequence of the finite
differencing and has no physical counterpart in the PDEs.
A modal stability analysis of finite difference equations derived
from the GHD PDEs does not yield the Jeans condition,
but rather a generalized Courant condition that
includes effects of gravity (\citeauthor{tru98} \citeyear{tru98}).


\section{APPLICATIONS TO ROTATING MOLECULAR CLOUDS}

In this section we briefly describe recent high-resolution results we
have obtained
 with AMR
for the gravitational collapse of nonisothermal, rotating, uniformly
dense clouds (\citeauthor{fisher98} \citeyear {fisher98}).
Isothermal centrally condensed
clouds \cite{tru97} and isothermal uniform clouds \cite{tru98}
 have been previously discussed and we refer the reader to these papers. 

The uniform cloud is an idealized model of an astrophysical
cloud but is very useful as a first approximation. 
The initial isothermal configuration of the uniform cloud may be
completely parameterized by the dimensionless
energy ratios $\alpha\equiv\
E_{\rm thermal}/|E_{\rm gravitational}|$ and $\bo\equiv\
E_{\rm rotational}/|E_{\rm gravitational}|$.

\citeauthor{bur93} \citeyear{bur93}  
studied a uniform cloud, with
$M=1M_{\odot}$ and $R=5\times10^{16}\cm$,
which give $\rho_0=10^{-17.4}\gcc$.  Its
energy ratios are
$\alpha=0.26$
and $\bo=0.16$.  The isothermal uniform cloud thus has $\cs=0.167\kms$
 and a rotation rate
$\Omega=7.2\times10^{-13}\rads$.  The cloud is perturbed
by $\rho\rightarrow\rho\times[1+0.1\cos(2\phi)]$, a 10\% $m=2$ mode seed
 perturbation.
This cloud begins with
$\jl=1.17R$.  
We have used an initial resolution of $R_{32}$ and dynamically refined
so as to ensure $\jmax=0.25$.
In the initial configuration, the fiducial radius beyond which cloud gravity
dominates perturbation pressure is $\rcrit=0.14R$, so that
with $R_{32}$ we still linearly
resolve this scale with more than 4 cells. 

\subsection{Non-Isothermal Uniform Clouds}

In recent work using high resolution AMR
(\citeauthor{fisher98} \citeyear{fisher98}),   
we followed the collapse of an initially rigidly
rotating, uniform isothermal
cloud.  We used a two-component barotropic equation
of state which makes the transition from isothermal to 
polytropic in a smooth fashion. 
The initial conditions are identical to the
\citeauthor {bur93} \citeyear{bur93} isothermal uniform cloud perturbed
by a 10\% $m=2$ mode.
In the following we discuss the
evolution of this nonisothermal collapse.
While a similar calculation has been recently studied
by \citeauthor{bat97} (\citeyear{bat97}), to our knowledge, 
none of the calculations in
the literature to date have been able to follow the subsequent 
evolution of the fragments over dynamical (orbital) timescales, while
still adhering to the Jeans criterion. However, to be fully certain
of the reality of the fragmentation, it is still necessary to do a 
convergence study by decreasing the Jeans number (\citeauthor {tru97} 
\citeyear {tru97}), which we have not yet done for this calculation.

Figure 1 represents
a 2-D slice of log density through the equatorial plane 
$t=1.41\times 10^{12}$ s after start of the
collapse.
The cloud has initially collapsed to an isothermal 
disk, and an elongated filamentary bar with the first signs of
fragmentation apparent.  A strong isothermal shock
above the plane of the disk is soon established.  After $t=1.46\times 10^{12}$ s
(Figure 2)
 the isothermal bar becomes optically thick, and the accretion flow onto the bar
is arrested, resulting in the growth of non-axisymmetric perturbations
in the bar.
Fragmentation in the bar results in the formation of binary  spherical cores
 connected by a prominent bar.  
The core-bar system is embeddded in a two-armed spiral, derivative of
the original
$m=2$  perturbation.
The binary 
separation decreases as the
cores increase their mass by direct 
accretion from the low-angular momentum material
of the bar, eventually leading to
the dissipation of the bar.
Figure 3 shows a snapshot of log density  about
0.5 rotation periods later at $t=1.51\times 10^{12}$ s.  Protostellar disks
have now formed around the cores with the
 cores at their closest orbital separation $\sim 44$ AU.  The disks accrete gas
directly from the long spiral arms. 
Each core has $0.08 M_{\odot}$ with a radius $\sim 10$ AU.
The masses and radii of these
first collapse cores are in good agreement with the detailed
 1-D radiation-hydrodynamic calculations of \citeauthor{masu98} (\citeyear{masu98}).  The core accretion
luminosity  $\sim 2\times 10^{31}
\rm {erg s^{-1}}$ obtained in the 3-D collapse calculations is considerably
less than that found in the 1-D simulations,
$\sim 4\times 10^{32} \rm {erg s^{-1}}$, due to the
large angular momentum
barrier in the 3-D simulations, resulting in slower accretion onto the
protostellar cores. 

It is important to point out that the initial rapid growth of the cores is due
to accretion of matter from the bar.  As the bar dissipates, the cores have
ended their first phase of growth and proceed to grow more slowly
by direct accretion from the surrounding protostellar disks.  At $t=1.52\times 10^{12}$ s
(Figure 4) the cores begin to separate and the surrounding disks become morphologically
 distinct.
The protostellar disks and cores become a fully detached binary by $t=1.56\times 10^{12}$ s
(Figure 5) with the disks attached to the long spiral.  The scale
of the disks, of order 100 AU, is consistent with observations
of gaseous disks surrounding single T-Tauri stars (\citeauthor {sar94}
\citeyear {sar94}) and
debris disks surrounding systems such as $\beta$ Pictoris (\citeauthor {art97}
\citeyear {art97}).  When we stopped our
calculations, at $t=1.6\times
10^{12}$ s(Figure 6) the binary protostellar disks/core system has moved back into the
surrounding spiral arms.  The cores have 20\% of the mass of the initial cloud 
and the arms have 27\% of mass of the initial cloud at this time.  The protostellar disks
comprise about 2\% of the cloud mass.   

\section{CONCLUSIONS}

In this paper we have described a powerful new method for numerical
solution of 3-D self-gravitational hydrodynamic problems.  This method
 combines a Godunov
hydrodynamics integrator with a multigrid gravity solver in an
adaptive mesh refinement framework.  Guided by the Jeans condition
for isothermal problems, and the use of Richardson extrapolation,
AMR efficiently expends computational resources only when and
where the
features of the flow demand it.  We presented results of this
methodology applied to the collapse and fragmentation
of nonisothermal molecular clouds that are both initially uniform and
 centrally condensed,
evolved over 8--9 decades of collapse and subject to initial $m=2$ mode
perturbations.  We found binary fragmentation
results in good agreement with published
calculations after evolution over the first half of this
logarithmic range.  By automatically
inserting ever finer grids to maintain
resolution of the Jeans length, these calculations did not generate the
substantial artificial viscosity found in published
calculations that used fixed finest resolutions.
We have shown the first
well-resolved formation sequence of
protostellar disks surrounding a pair of binary cores, and discussed the 
role of the bar connecting the binary pair in the accretion onto the
cores.  The disks formed are consistent with observations of
gaseous disks surrounding single T-Tauri stars and debris disks
surrounding systems such as $\beta$ Pictoris.


\acknowledgements{
Research on star formation for RIK and CFM is 
supported  by a grant from NASA's Astrophysics Theory Program
to the Center for Star Formation Studies.
RIK acknowledges additional support under the auspices of the
US Department of Energy at the Lawrence Livermore National Laboratory
under contract W-7405-Eng-48. CFM's research is supported in part by NSF
grant AST9530480. We wish to thank the Pittsburgh
Supercomputing Center for provision of Cray C90 resources through
grant AST940011P.}



\begin{thebibliography}{}
\bibitem[\protect\citeauthoryear{Artymowicz}{1997}]{art97} Artymowicz, P. 1997, Ann. Rev. Planet. Sci., 25, 175
\bibitem[\protect\citeauthoryear{Bate\&Burkert}{1997}]{bat97} Bate, M.R.\& Burkert, A. 1997, M.N.R.A.S., 288, 1060
\bibitem[\protect\citeauthoryear{Bell, Colella, \& Trangenstein}{1989}]{bel89} Bell, J.B., Colella, P., \&
 Trangenstein, J.  1989, Journal of Computational Physics, 82, 362
\bibitem[\protect\citeauthoryear{Berger \& Colella}{1989}]{ber89} Berger, M.J. \& Colella, P.
  1989, Journal of Computational Physics, 82, 64
\bibitem[\protect\citeauthoryear{Berger \& Oliger}{1984}]{ber84} Berger, M.J. \& Oliger, J.
  1984, Journal of Computational Physics, 53, 484
\bibitem[\protect\citeauthoryear{Boss}{1991}]{bos91} Boss, A.P.  1991, Nature, 351, 298
\bibitem[\protect\citeauhtoryear{Boss}{1996}]{bos96} Boss, A.P.  1996, ApJ, 468, 231
\bibitem[\protect\citeauthoryear{Burkert \& Bodenehimer}{1993}]{bur93} Burkert, A. \&
 Bodenheimer, P.  1993, M.N.R.A.S., 264, 798
\bibitem[\protect\citeauthoryear{Colella, \& Glaz}{1985}]{col85} Colella, P. \& Glaz, H.M.  1985,
 Journal of Computational Physics, 59, 264
\bibitem[\protect\citeauthoryear{Colella \& Woodward}{1984}]{col84} Colella, P. \& Woodward, P.R.
 1984, Journal of Computational Physics, 54, 174
\bibitem[\protect\citeauthoryear{Fisher, Klein, \& McKee}{1998}]{fisher98} Fisher, R., Klein, R., \& McKee,
 C. 1998, in preparation
\bibitem[\protect\citeauthoryear{Inutsuka \& Miyama}{1992}]{inu92} Inutsuka, S. \& Miyama, S.M.
 1992, ApJ, 388, 392
\bibitem[\protect\citeauthoryear{Inutsuka, \& Miyama}{1997}]{inu97} Inutsuka, S. \& Miyama, S.M.
 1997, ApJ, 480, 681
\bibitem[\protect\citeauthoryear{Jeans}{1902}]{jea02} Jeans, J.H.  1902, Phil. Trans. A, 199, 1
\bibitem[\protect\citeauthoryear{Jeans}{1928}]{jea28} Jeans, J.H.  1928, Astronomy \& Cosmogony,
 (London: Cambridge University Press)
1993, A\&A, 273, 175
\bibitem[\protect\citeauthoryear{Klein, \& McKee}{1994}]{kle94a} Klein, R.I. \& McKee, C.F.  1994,
 in Numerical Simulations in Astrophysics, ed. J. Franco et al.
 (New York: Cambridge), 251
\bibitem[\protect\citeauthoryear{Klein, McKee, \& Colella}{1994}]{kle94b} Klein, R.I., McKee, C.F., \&
 Colella, P.  1994, ApJ, 420, 213
\bibitem[\protect\citeauthoryear{Klein, McKee, \& Woods}{1995}]{kle95} Klein, R.I., McKee, C.F., \&
 Woods, D.T.  1995, in Physics of the Interstellar Medium and
 Intergalactic Medium, ed. A. Ferrara et al. (San Francisco:
 Astronomical Society of the Pacific), 366
\bibitem[\protect\citeauthoryear{Klein \& Woods}{1998}]{kle98} Klein, R.I. \& Woods, D.T.  1998, ApJ, 497,777
\bibitem[\protect\citeauthoryear{Klein}{1998}]{klein98} Klein, R.I. 1998,
Journal of Computational and Applied Mathematics, ed. H. Riffert (Elsevier Press), in press
\bibitem[\protect\citeauthoryear{LeVeque}{1992}]{lev92} LeVeque, R.J.  1992, Numerical Methods for
 Conservation Laws, (Boston: Birkhauser Verlag)
\bibitem[\protect\citeauthoryear{Masunaga, Miyama, \& Inutsuka}{1998}]{masu98} Masunaga, H., Miyama, S. \& Inutsuka, S.
1998, ApJ, 495, 346
\bibitem[\protect\citeauthoryear{Richtmyer \& Morton}{1967}]{ric67} Richtmyer, R.D. \& Morton,
 K.W. 1967, Difference Methods for Initial-Value Problems,
 (New York: Interscience)
\bibitem[\protect\citeauthoryear{Sargent \& Beckwith}{1994}]{sar94} Sargent, A.I, \& Beckwith, S.V.W., 1994, Astroph. and Space Sci., 212, 181.
\bibitem[\protect\citeauthoryear{Truelove {\it et al.}}{1997}]{tru97} Truelove, J.K., 
 Klein, R.I., McKee, C.F., Holliman, J.H., II, Howell,
 L.H., \& Greenough, J.A. 1997, ApJ, 489L,179
\bibitem[\protect\citeauthoryear{Truelove {\it et al.}}{1998}]{tru98} Truelove, J.K., 
 Klein, R.I., McKee, C.F., Holliman, J.H., II, Howell,
 L.H., Greenough, J.A., \& Woods, D.T. 1998, ApJ,495,821
\bibitem[\protect\citeauthoryear{Winkler \& Newman}{1980}]{win80} Winkler, K.-H.A. \& Newmanm, M.J. 1980,
 ApJ, 238, 311
\bibitem[\protect\citeauthoryear{Woods {\it et al.}}{1996}]{woo96} Woods, D.T., Klein, R.I., Castor,
 J.I., McKee, C.F., \& Bell, J.  1996, ApJ, 461, 767
\end{thebibliography}
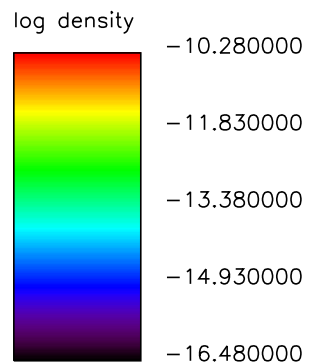
\end{document}